\documentclass[10pt,a4paper]{article}
\usepackage{graphicx}
\usepackage{epstopdf}
\usepackage[margin=0.8in,top=1.2in]{geometry}

	\usepackage{fancyheadings}
\renewcommand{\thispagestyle}[1]{} % do nothing

\begin{document}

\title{A simple thermodynamic description of the combined Einstein and elastic models}

\author{Tadeusz Balcerzak, Karol Sza{\l}owski\thanks{E-mail: \texttt{kszalowski@uni.lodz.pl}}\\ \\
  {\normalsize Department of Solid State Physics,}\\ {\normalsize University of \L\'{o}d\'{z},}\\ {\normalsize Pomorska 149/153, 90-236 \L\'{o}d\'{z}, Poland}\\
   \\ Michal Ja\v{s}\v{c}ur \\ \\{\normalsize Department of Theoretical Physics and Astrophysics, Faculty of Science,}\\ {\normalsize P. J. \v{S}afarik University,}\\ {\normalsize Park Angelinum 9, 041 54 Ko\v{s}ice, Slovak Republic}}

%\author{Tadeusz Balcerzak\thanks{Department of Solid State Physics, University of \L\'{o}d\'{z}, Pomorska 149/153, 90-236 %\L\'{o}d\'{z}, Poland}, Karol Sza{\l}owski$^{*}$ and Michal Ja\v{s}\v{c}ur\thanks{Department of Theoretical Physics and %Astrophysics, Faculty of Science, P. J. \v{S}afarik University, Park Angelinum 9, 041 54 Ko\v{s}ice, Slovak Republic}}

\date{25 July 2010}
%\ead{kszalowski@uni.lodz.pl}

\maketitle

\begin{abstract}
Simple application of the Einstein model combined with the elastic description of solid state is developed. The frequency of quantum oscillators has been assumed as volume dependent and, furthermore, elastic energy terms of static character have been included to complete the description. Such an extension enables to construct the complete thermodynamics. In particular, the model yields practical equation of state and describes the thermal expansion coefficient as well as the isothermal compressibility of solids. The thermodynamic properties resulting from the Gibbs free-energy analysis have been calculated and illustrated in figures. Some comparison of the theoretical results with experimental data for solid argon has been made.
\end{abstract}

%\pacs{63.70.+h, 64.10.+h, 65.50.+m}

%\submitto{\JPCM}

\section{Introduction}

The Einstein model of the solid state has been known for many years as the first model able to describe qualitatively the low-temperature behaviour of the specific heat \cite{einstein,kittel,girifalco}. This model often serves as an approximation for studies of optical phonons, or soft modes in some intermetallic compounds \cite{caplin} and, as well, is found suitable to describe thermal properties of some modern low-dimensional structures \cite{avramov}. It is also referred to as a prototype for which the more sophisticated Debye model has been developed \cite{debye}. Nonetheless, it has been shown that in many cases the Einstein model provides better results than the Debye model \cite{zubov}. It is known that in spite of its usefulness and simplicity, the pure Einstein model lacks the ability to describe fully the thermodynamic properties of the solid state. For instance, neither the proper equation of state, nor the thermal expansion or compressibility can be obtained. Thus, apart from the celebrated specific heat behaviour, which is an important consequence of the quantum nature of harmonic oscillators, other thermodynamic properties have not successfully been described within this model.\\

There are numerous attempts in the literature aimed at improving the pure Einstein model \cite{li, cankurtaran, lusk, holzapfel, ponkratz}.
For instance, a "variational Einstein model" has been developed in Ref.~\cite{li} for describing low temperature solids from the Feynman path integral perspective. The formalism has been applied to a specific system, consisting of solid hydrogen with lithium impurities. However, one of the theoretical results of Ref.~\cite{li} seems to be controversial, namely the dependence of the free energy, which is presented there as an increasing function of temperature (see Fig.16 of Ref.~\cite{li}), giving the indication that the corresponding entropy is negative. 
Cankurtaran and Askerov \cite{cankurtaran} have introduced the so called Einstein-Debye model for which the calculations of the thermal expansion coefficient were possible by introducing the Gr\"uneisen parameter.
However, neither the Gibbs free-energy, nor the compressibility could be calculated within that approach. Thus, the complete thermodynamic description has not been achieved there. 
In Ref.~\cite{lusk} the "nonlocal Einstein crystal" has been considered, for which the thermodynamic functions have been constructed. The considerations presented in Ref.~\cite{lusk} seem to be of pure theoretical character when the Einstein crystal is replaced by a harmonic sublattice with a size that is coupled to the total volume.
Additionally, the Mie-Gr\"uneisen approach has been developed by Holzapfel et al. for the description of simple metals \cite{holzapfel} as well as some simple solids, such as NaCl and MgO \cite{ponkratz}. The optimized pseudo-Debye-Einstein model has been used there, and good agreement with the experimental data has been achieved. However, the method presented in Refs.~\cite{holzapfel, ponkratz} does not seem to be straightforward enough to use since it requires many fitting parameters to be introduced for calculations of the free-energy.\\

On the other hand, there exists a rich literature on the description of solid state properties within the elastic models \cite{murnaghan, birch1, slater, bridgman, barsch, wallace1, birch2, vinet2, vinet, wallace2}.
These studies are supported by the experimental measurements \cite{anderson1, anderson2}.
For instance, Birch \cite{birch2} considers the single-crystal and polycrystalline NaCl at high pressure using the finite strain (Eulerian) theory. In this approach, the equation of state contains coefficients which have to be fitted in each particular temperature. Despite the usefulness of the model for studies of isothermal compression, it does not involve the thermal (vibrational) energy and therefore is not suitable for description of the specific heat. Similarly, in the paper of Vinet et al. \cite{vinet2,vinet} the universal form of the equation of state has been found based on the scaling argumentation. The results obtained there have been discussed in the context of the Birch-Murnaghan equation of state \cite{wallace1} and show very good agrement with the experimental data. However, the Gibbs free-energy has not been derived in that works (Ref.~\cite{vinet2,vinet}), so that the full thermodynamic description of the system has not been constructed. As a matter of fact, the same deficiency can be risen for some other papers devoted to the application of the equation of state based on the elastic models. A comparative review of the experimental equations of state existing in the literature is given in Refs.~\cite{roy,hama}.\\

Among other modern methods one should mention the first principles computations within the density-functional theory (DFT). This includes, for instance, the equation of state, elastic constants and phonon dispersion relation calculations \cite{foata}. On the other hand, by the first-principles molecular-dynamics simulation (FPMD) method the thermal expansivity and the specific heat have been calculated \cite{wu}. A good agreement with the experimental data is often achieved. However, the analytic description remains always desirable from the point of view of understanding the physics behind the model.\\

Such a situation motivated us to combine the Einstein and elastic  models and to complete them in a way that enables to overcome  the above-mentioned problems. Looking in the literature, such idea can already be found in some papers \cite{chisolm, greeff}, however, only particular thermodynamic characteristics have been studied there. Thus, our aim is to construct the self-consistent and full thermodynamic description. In order to make it possible, the following  aspects should be taken into account: Firstly, the common frequency of quantum oscillators should not be treated as a constant value but should be volume (and, in consequence, temperature) dependent. Obviously, such idea is not new and for the first time emerged in the paper of Gr\"uneisen \cite{gruneisen}, where a specific  frequency/volume dependence has been assumed as a hypothesis. This assumption is supported by the argument that the wavelength of collective excitations (phonons) depends on the crystal size. 
Secondly, the vibrational Einstein Hamiltonian necessarily needs to be completed by the static, elastic part \cite{kwon}. The elastic energy accounts for the mutual interactions of the atoms even if they are not in a thermal vibrating state \cite{wallace1, wallace2}.
It is known that the static energy is responsible for crystal compressibility \cite{wallace1}, and owing to its magnitude and nature, cannot be inferred from the model of independent oscillators. It turns out that for the purpose of the combined model the elastic free-energy should also be modified by taking into account the linear term (which is normally neglected in the elastic theory) as well as the static entropy. The details of this modification will be explained in the next section. Owing to analytical simplicity we will neglect the electronic excitation term. It is argued that this term is generally a small correction to the solid equation of state \cite{greeff}.\\

Taking this into account, we propose  improving the model and constructing the Gibbs energy of the system from the beginning. We assume that the balance between the internal and external pressures keeps the system in mechanical equilibrium. 
According to our knowledge, the detailed balance between the expanding pressure of quantum oscillators and the compressive elastic pressure has not been discussed in the literature.
Having obtained the expression for the Gibbs energy, a full thermodynamic description of the crystal can be achieved.
The main parameters of the theory, which can be extracted from experiment, are the Einstein frequency in the ground state (or the Einstein temperature), the volume elastic modulus in the ground state - supplemented by the structural space-filling coefficient - and the Gr\"uneisen parameter. We will show that these parameters, in principle, allow us to calculate other properties within a reasonable range of experimental values.\\

The paper is organized as follows: In the following, theoretical section, the outline of the model is  given and the method of derivation the Gibbs energy is presented. In particular, for the combined model the new equation of state is obtained. In the last section, some representative numerical results concerning the thermodynamic properties are illustrated in figures. These calculations are compared with the experimental data for solid argon. A critical discussion of the presented approach is also included.\\

%section 2
\section{Theoretical model}

\subsection{The model Hamiltonian}

The Hamiltonian is assumed in the form of:
\begin{equation}
\label{eq1}	
\mathcal{H}=\mathcal{H}_{\varepsilon} + \mathcal{H}_{\omega},
\end{equation}
where the elastic, volume dependent part, $\mathcal{H}_{\varepsilon}$, can be written as:
\begin{equation}
\label{eq2}	
\mathcal{H}_{\varepsilon}=NA\,\varepsilon+\frac{1}{2}NB\,\varepsilon^2 - \frac{1}{ 3!}NC\,\varepsilon^3 + \frac{1}{ 4!}ND\,\varepsilon^4,
\end{equation}
whereas the oscillatory part, $\mathcal{H}_{\omega}$, is given in the form:
\begin{equation}
\label{eq3}	
\mathcal{H}_{\omega}=\sum_{i=1}^{3N}{\hbar \omega \left(\hat{n}_i+\frac{1}{2}\right)}.
\end{equation}
In (\ref{eq2}) and (\ref{eq3}) $N$ is the number of atoms which are treated as the three-dimensional oscillators, and $\hat{n}_i$ is the excitation number operator connected with the $i$-th oscillator. 
The form of the elastic Hamiltonian (\ref{eq2}) assumed in this paper is different (much simpler) from that presented in 
Refs.~\cite{chisolm} or \cite{greeff}.
The elastic Hamiltonian, $\mathcal{H}_{\varepsilon}$, consists of several terms. The most important is the harmonic term $\left(\propto \varepsilon^2\right)$, where $B$-constant is the volume elastic modulus in the ground state. We found that the linear term  $\left(\propto \varepsilon \right)$ is also necessary, where $A$-constant is responsible for the internal (compressive) pressure. Although the linear, anharmonic part is assumed to be very small in comparison to the harmonic one ($A\ll B$), its role is very important in balancing the internal (expanding) pressure produced by the anharmonic oscillators, i.e., by the $\mathcal{H}_{\omega}$ part. A certain requirement for $A$-values is imposed as the equilibrium condition for the whole energy, which is discussed later. The relative elastic deformation $\varepsilon$ is defined by the relation:
\begin{equation}
\label{eq4}	
V=V_0\left(1+\varepsilon \right),
\end{equation}
where $V$ is the crystal volume in a current thermodynamic state, and $V_0$ is the volume at external pressure $p=0$ (vacuum) and temperature $T=0$. Equation (\ref{eq4}) indicates that a non-zero value of volume deformation $\varepsilon$ occurs when the external pressure $p \ne 0$ and/or the temperature $T>0$ is applied. Thus, a possibility of introducing the isothermal compressibility and thermal expansion coefficient has been opened within the presented approach.\\

A composition of the Hamiltonian (\ref{eq1}) consisting of the classical ($\mathcal{H}_{\varepsilon}$) and quantum ($\mathcal{H}_{\omega}$) parts is justified for the purpose of constructing the thermodynamics \cite{wallace2}.  Obtaining the free energy of the full system is one of the main goals and that can be conveniently done by summing up the free energies corresponding to those two parts. A similar modus operandi has also been presented in the paper Ref.~ \cite{kwon} albeit for a different approach. It should be stressed that by introducing the elastic Hamiltonian, $\mathcal{H}_{\varepsilon}$, the static non-vibrational energy has been taken into account, which is crucial for the equation of state. It is worth mentioning that the method can be further generalized by introducing also the electronic part of the free energy \cite{wallace2}.

\subsection{The elastic free-energy}

In order to calculate the elastic (Helmholtz) free energy, one has to evaluate the static internal energy and the static entropy.  The static internal energy $U_\varepsilon$ can be immediately found from the classical part of the Hamiltonian:
\begin{equation}
\label{eq5}	
U_{\varepsilon} = \mathcal{H}_{\varepsilon}
\end{equation}
From Eq.(\ref{eq2}) it is seen that this energy is explicitly volume-dependent. On the other hand, the static entropy $S_{\varepsilon}$ is connected with filling up the volume of the system in the absence of thermal movement. Let us denote by $V_{\mathrm ex}$ the own volume occupied by $N$ atoms in the crystal state. For a given crystallographic lattice we can introduce the space filling coefficient $q=V_{\mathrm ex}/V_0$. For instance, for FCC and HCP structures and the model of hard spheres $q \approx 0.74$. Thus, inside each primitive cell $q= \rho_1$ is the probability of finding an atom, whereas $1-q= \rho_2$ is the probability of opposite event. The static entropy connected with the bimodal distribution $\rho_i$  (where $i=1,2$) can be calculated from the mean value of $\ln \rho_i$, using the general formula for the exact differential of entropy: \cite{hill}
\begin{equation}
\label{eq6a}	
dS_{\varepsilon} = -N k_{\mathrm B} d\langle \ln \rho_i \rangle.
\end{equation}
Hence, in our case:
\begin{equation}
\label{eq6}	
S_{\varepsilon} = -N k_{\mathrm B} \left[q\ln q + \left(1-q\right) \ln \left(1-q \right) \right] -N k_{\mathrm B}c
\end{equation}
where $c$ is a constant of integration. Since $c$ is arbitrary, as a matter of convenience we can choose $c=0$. This arbitrariness is due to the fact that only changes in entropy are experimentally measurable and have thermodynamic significance \cite{hill,reif}.\\

The entropy is additive with respect to the number of primitive cells $N$. For the purpose of this paper it is assumed in the first approximation that the space-filling coefficient $q$ is volume independent and, in consequence, the corresponding entropy $S_{\varepsilon}$ is constant. Thus, the static entropy (for $c=0$) results in the residual entropy of the crystal for $T \to 0$, which is due to the fact that the whole volume is not perfectly filled. Moreover, the important role of $S_{\varepsilon}$ 
(being dependent on $q$)
may manifest itself in the structural (1st order) phase transitions \cite{wallace2}.\\

The elastic, Helmholtz free-energy can now be found from the formula:
\begin{equation}
\label{eq7}	
F_{\varepsilon} = U_{\varepsilon}-T S_{\varepsilon}.
\end{equation}
Hence, we finally obtain:
\begin{eqnarray}
\label{eq8}	
F_{\varepsilon} &=& NA\,\varepsilon+\frac{1}{2}NB\,\varepsilon^2 - \frac{1}{ 3!}NC\,\varepsilon^3 + \frac{1}{ 4!}ND\,\varepsilon^4\nonumber\\
& +& N k_{\mathrm B} T\left[q\ln q + \left(1-q\right) \ln \left(1-q \right) \right] .
\end{eqnarray}
For the full thermodynamic description this energy should be completed by the vibrational, Einstein part.

\subsection{The vibrational free-energy}

For the Hamiltonian (\ref{eq3}) it is convenient to employ the canonical ensemble. The Helmholtz free-energy can be found from the formula:
\begin{equation}
\label{eq9}	
F_{\omega} = - k_{\mathrm B} T \ln Z_{\omega}
\end{equation}
where the statistical sum $Z_{\omega}$ can be calculated exactly as:
\begin{equation}
\label{eq10}	
Z_{\omega} = {\mathrm Tr}\,e^{- \beta \mathcal{H}_{\omega}} = \left[2 \sinh \left(\frac{1}{2}\beta \hbar \omega \right) \right]^{-3N}
\end{equation}
where $\beta=1/k_{\mathrm B} T$.
Thus we obtain:
\begin{equation}
\label{eq11}	
F_{\omega} = 3N k_{\mathrm B} T \ln \left[2 \sinh \left(\frac{1}{2}\beta \hbar \omega \right) \right]
\end{equation}
On the other hand, the vibrational internal energy $U_{\omega}$ is given by the statistical mean value of the corresponding Hamiltonian:
\begin{equation}
\label{eq12}	
U_{\omega} = \left<\mathcal{H}_{\omega}\right> = 3N \hbar \omega \left(\left< \hat{n}_i \right> +\frac{1}{2}\right) = \frac {3}{2} N \frac {\hbar \omega} {\tanh \left(\frac{1}{2}\beta \hbar \omega \right)}.
\end{equation}
Hence, the vibrational entropy $S_{\omega}$ can be found from the relationship:
\begin{eqnarray}
\label{eq13}	
S_{\omega} &=& \frac {U_{\omega}-F_{\omega}}{T} \nonumber\\
&=& \frac {3N}{2T}\frac {\hbar \omega} {\tanh \left(\frac{1}{2}\beta \hbar \omega \right)}-3Nk_{\mathrm B} \ln \left[2 \sinh \left(\frac{1}{2}\beta \hbar \omega \right) \right].
\end{eqnarray}
The specific heat at constant $V$ is given by the well-known formula:
\begin{eqnarray}
\label{eq14}
C_{V} &=& T \left( \frac{\partial S_{\omega}}{\partial T}\right)_V = 
\left( \frac{\partial U_{\omega}}{\partial T}\right)_V \nonumber\\
&=&3N k_{\mathrm B} \left(\frac {\Theta}{2T}\right)^2 \sinh ^{-2} \left(\frac {\Theta}{2T}\right)
\end{eqnarray}
where $\Theta = \hbar \omega /k_{\mathrm B}$.\\
Further, in order to improve the Einstein model it is assumed that $\omega$ is not a constant but volume dependent according to the Gr\"uneisen assumption \cite{gruneisen}:
\begin{equation}
\label{eq15}	
\omega \propto \frac{1}{V^{\gamma}} 
\end{equation}
where $\gamma$ is the Gr\"uneisen constant. Thus, we can write:
\begin{equation}
\label{eq16}	
\omega = \frac{\omega_0}{\left(1+\varepsilon \right)^{\gamma}},
\end{equation}
where we made use of the relation (\ref{eq4}), and $\omega_0$ in (\ref{eq16}) is the Einstein frequency in the ground state (i.e., at $p=0$, $T=0$, and $\varepsilon =0$). Then, the variable $\Theta = \hbar \omega /k_{\mathrm B}$  can be presented in the form of:
\begin{equation}
\label{eq17}	
\Theta = \frac{\Theta _0}{\left(1+\varepsilon \right)^{\gamma}}
\end{equation}
where
\begin{equation}
\label{eq18}	
\Theta _0= \frac{\hbar \omega _0}{k_{\mathrm B}}.
\end{equation}
$\Theta _0$ is the so-called Einstein temperature. The correction of the Einstein model presented by Eq.(\ref{eq17}) implies that all the thermodynamic potentials will depend on the volume via elastic deformation $\varepsilon$.

\subsection{Thermodynamic functions of the combined model}

The total Helmholtz free-energy is given by the sum of expressions (\ref{eq8})  and  (\ref{eq11}), with the help of (\ref{eq15})-(\ref{eq18}):
\begin{eqnarray}
\label{eq19}	
F &=& NA\,\varepsilon+\frac{1}{2}NB\,\varepsilon^2 -\frac{1}{ 3!}NC\,\varepsilon^3 +\frac {1}{ 4!}ND\,\varepsilon^4 \nonumber\\
&+& N k_{\mathrm B} T\left[q\ln q + \left(1-q\right) \ln \left(1-q \right) 
\right]\nonumber\\  
&+& 3N k_{\mathrm B} T \ln \{ 2 \sinh \left[\frac{\Theta _0}{2 T} \frac{1}{\left( 1+ \varepsilon \right)^{\gamma}} \right] \}.
\end{eqnarray}
The total Gibbs energy is then given by:
\begin{equation}
\label{eq20}
G(p,T) = F + pV
\end{equation}
The equation of state can be obtained from the variational principle:
\begin{equation}
\label{eq21}
\left( \frac{\partial G}{\partial \varepsilon}\right)_{p,T} =0 
\end{equation}
which, from (\ref{eq20}), is equivalent to:
\begin{equation}
\label{eq22}
p V_0 = -\left( \frac{\partial F}{\partial \varepsilon}\right)_T 
\end{equation}
where $F$ is given by (\ref{eq19}). From (\ref{eq22}), after differentiation of (\ref{eq19}), we obtain:
\begin{eqnarray}
\label{eq23}
p V_0 &=& -NA -NB \varepsilon + \frac{1}{2}NC {\varepsilon}^2 
- \frac{1}{3!}ND {\varepsilon}^3 \nonumber\\
&+&\frac {3}{2}\gamma \frac{N k_{\mathrm B} \Theta_0} {\left( 1+ \varepsilon \right)^{\gamma +1} \tanh \left[\frac {\Theta_0}{2 T} \frac{1}{\left( 1+ \varepsilon \right)^{\gamma}} \right]}.
\end{eqnarray}
For the high-temperature limit equation (\ref{eq23}) takes a form:
\begin{equation}
\label{eq23a}
p=\,-\frac{\partial F_{\varepsilon}}{\partial V}+\frac{3 N k_{\mathrm B} T}{V}\gamma
\end{equation}
which is the classical lattice dynamics pressure \cite{wallace2}.
On the other hand, in the low-temperature limit ($T \to 0$) and for small pressure the deformation $\varepsilon$ is small too, and Eq.(\ref{eq23}) can be linearized as:
\begin{equation}
\label{eq24}
p V_0 \approx -NA -NB \varepsilon+
\frac{3}{2}\gamma N k_{\mathrm B} \Theta_0 \left[1-\left(\gamma+1\right)\varepsilon \right].
\end{equation}
Naturally,  when $p=0$ then in the ground state is  $\varepsilon =0$ as well. Consequently,  we have to demand that the internal pressure of oscillators, which then is equal to $\frac{3}{2}\gamma N k_{\mathrm B} \Theta_0$, cancels out the pressure arising from the anharmonic, linear term, i.e., $-NA$. Then the system is in equilibrium without any external force. The condition for this demand leads to the specific choice for the anharmonic constant $A$, namely:
\begin{equation}
\label{eq25}
A = 3 \gamma A_0
\end{equation}
where
\begin{equation}
\label{eq26}	
A_0= \frac{1}{2} \hbar \omega _0 = \frac{1}{2}k_{\mathrm B} \Theta_0.
\end{equation}
The energy constant $A_0$ presents a convenient unit for introduction of the dimensionless external pressure $\pi$ defined by:
\begin{equation}
\label{eq27}
\pi = \frac{V_0}{N A_0}\,p
\end{equation}
and the dimensionless temperature $\tau$:
\begin{equation}
\label{eq28}
\tau = \frac{k_{\mathrm B} T}{A_0}.
\end{equation}
With the above quantities, and Eqs.(\ref{eq25}) - (\ref{eq26}), the equation of state (\ref{eq23}) takes the simple form:
\begin{eqnarray}
\label{eq29}
\pi +3 \gamma + \frac{B}{A_0}\varepsilon &-& \frac{1}{2}\frac{C}{A_0}\varepsilon^2 + \frac{1}{3!}\frac{D}{A_0}\varepsilon^3\nonumber\\
&=& 3 \gamma \frac {1}{\left(1+ \varepsilon \right)^{\gamma +1}} \tanh ^{-1} \left[\frac{1}{\tau \left(1+ \varepsilon \right)^{\gamma}}\right].
\end{eqnarray}
Eq.(\ref{eq29}), obtained within the approximations assumed in this paper, presents a dimensionless equation of state from which the elastic deformation $\varepsilon (\pi,\tau)$ can be calculated for arbitrary temperature $\tau$ and external pressure $\pi$.\\
The Gibbs free-energy (\ref{eq20}) in the dimensionless form can be written as follows:
\begin{eqnarray}
\label{eq30}
\frac{G(p,T)}{N A_0} &=& 3 \gamma\varepsilon+ \frac{1}{2}\frac{B}{A_0}\varepsilon^2 - \frac{1}{3!}\frac{C}{A_0}\varepsilon^3 + \frac{1}{4!}\frac{D}{A_0}\varepsilon^4 \nonumber\\
&+& \tau \left[q\ln q + \left(1-q\right) \ln \left(1-q \right) 
\right]\nonumber\\  
&+& 3\tau \ln \{ 2 \sinh \left[\frac{1}{\tau \left( 1+ \varepsilon \right)^{\gamma}} \right] \}
+ \pi \left(1+\varepsilon \right),
\end{eqnarray}
where $\varepsilon = \varepsilon \left( \pi,\tau \right)$ is given by (\ref{eq29}). 
It is easy to show that from the formula (\ref{eq30}) all thermodynamic properties can be derived self-consistently. For instance, the total entropy fulfills the relationship:
\begin{equation}
\label{eq31}
S = - \left( \frac{\partial G}{\partial T}\right)_p = -\frac{k_{\mathrm B}}{A_0}\left( \frac{\partial G}{\partial \tau}\right)_{\pi}
\end{equation}
and, with the use of (\ref{eq29}), it can be obtained in the dimensionless form:
\begin{eqnarray}
\label{eq32}
\frac{S}{N k_{\mathrm B}}&=& - q\ln q - \left(1-q\right) \ln \left(1-q \right) \nonumber\\
&+& \frac {3}{\tau \left(1+ \varepsilon \right)^{\gamma}} \tanh ^{-1} \left[\frac{1}{\tau \left(1+ \varepsilon \right)^{\gamma}}\right]\nonumber\\
 &-& 3\ln \{ 2 \sinh \left[\frac{1}{\tau \left( 1+ \varepsilon \right)^{\gamma}} \right] \}.
\end{eqnarray}
On the other hand, the volume of the sample satisfies the equation:
\begin{equation}
\label{eq33}
V = \left( \frac{\partial G}{\partial p}\right)_{T} = 
\frac{V_0}{N A_0} \left( \frac{\partial G}{\partial \pi}\right)_{\tau}, 
\end{equation}
which again leads to the equation of state (\ref{eq29}). Other thermodynamic properties follow from the second derivatives of the Gibbs energy. For instance, the heat capacity at constant $p$ is given by:
\begin{equation}
\label{eq34}
C_p = -T\left( \frac{\partial^2 G}{\partial T^2}\right)_{p} = 
- \frac{k_{\mathrm B}}{A_0}\, \tau \left( \frac{\partial^2 G}{\partial \tau^2}\right)_{\pi}, 
\end{equation}
whereas the heat capacity at constant $V$ can be calculated from the relationship:
\begin{equation}
\label{eq35}
C_V = -T\left( \frac{\partial^2 F}{\partial T^2}\right)_{V} =
\left( \frac{\partial U}{\partial T}\right)_{V} = 
\frac{k_{\mathrm B}}{A_0} \left( \frac{\partial U}{\partial \tau}\right)_{\varepsilon}.
\end{equation}
On the other hand, the internal energy, $U$, can be written in the dimensionless form:
\begin{eqnarray}
\label{eq36}
\frac{U}{N A_0} &=& 3 \gamma\varepsilon+ \frac{1}{2}\frac{B}{A_0}\varepsilon^2 - \frac{1}{3!}\frac{C}{A_0}\varepsilon^3 + \frac{1}{4!}\frac{D}{A_0}\varepsilon^4  \nonumber\\
&+&3 \frac {1}{\left(1+ \varepsilon \right)^{\gamma}} \tanh ^{-1} \left[\frac{1}{\tau \left(1+ \varepsilon \right)^{\gamma}}\right].
\end{eqnarray}
Application of (\ref{eq36}) in (\ref{eq35}) leads to the dimensionless expression for $C_V$:
\begin{equation}
\label{eq37}
\frac{C_{V}}{N k_{\mathrm B}} =
3 \left[\frac {1}{\tau \left(1+ \varepsilon \right)^{\gamma}}\right]^2 \sinh ^{-2} \left[\frac {1}{\tau \left(1+ \varepsilon \right)^{\gamma}} \right].
\end{equation}
It should be noted that for the particular case, when $\varepsilon=0$, Eq.(\ref{eq37}) is equivalent to formula
(\ref{eq14}). However, in general, a non-zero $\varepsilon$ should be determined from the equation of state (\ref{eq29}).\\
Among other response functions which can be calculated either from the Gibbs free energy, or from the equation of state is, for example, the thermal expansion coefficient:
\begin{equation}
\label{eq38}
\alpha _p = \frac{1}{V}\left( \frac{\partial V}{\partial T}\right)_{p}
= \frac{k_{\mathrm B}}{A_0}\,\frac{1}{1 + \varepsilon}\left( \frac{\partial \varepsilon}{\partial \tau}\right)_{\pi}, 
\end{equation}
and the isothermal compressibility:
\begin{equation}
\label{eq39}
\kappa _T = -\frac{1}{V}\left( \frac{\partial V}{\partial p}\right)_{T}
= -\frac{V_0}{N A_0}\, \frac{1}{1 + \varepsilon} \left( \frac{\partial \varepsilon}{\partial \pi}\right)_{\tau}. 
\end{equation}
In particular, for $p=0$ and $T=0$, i.e., in the ground state, these response functions can be obtained from the approximate equation of state (\ref{eq24}). For this purpose, Eq.(\ref{eq24}) can be re-written in the form:
\begin{equation}
\label{eq40}
\varepsilon \approx  - \frac{V_0}{N A_0}\,\frac{1}{3 \gamma\left(\gamma +1\right) + B/A_0}\,p.
\end{equation}
It can be observed that Eq.(\ref{eq40}) for $p<0$ presents Hooke's law at $T=0$. 
Further, we can formally make use of the linear expansion of volume, namely:
\begin{equation}
\label{eq41}
V(p,T) \approx V_0 + \left( \frac{\partial V}{\partial T}\right)_
{p = 0 \choose T = 0} T + \left( \frac{\partial V}{\partial p}\right)_
{p = 0 \choose T = 0} p.
\end{equation}
It can be noticed that Eq.(\ref{eq41}) can equivalently be written as:
\begin{equation}
\label{eq42}
\varepsilon \approx \alpha _0 \,T - \kappa _0 \,p
\end{equation}
where $\alpha_0$ and $\kappa_0$ are the response functions (\ref{eq38}) and (\ref{eq39}), respectively, and these functions are taken in the ground state.
Now, by comparison of Eqs.(\ref{eq40}) and (\ref{eq42}) we obtain the result:
\begin{equation}
\label{eq43}
\alpha _0 \stackrel{def}{=} \frac{1}{V_0}\left( \frac{\partial V}{\partial T}\right)_{p = 0 \choose T = 0} = 0
\end{equation}
and
\begin{equation}
\label{eq44}
\kappa _0 \stackrel{def}{=} -\frac{1}{V_0}\left( \frac{\partial V}{\partial p}\right)_
{p = 0 \choose T = 0} = \frac{V_0}{N A_0}\,\frac{1}{3 \gamma \left(\gamma +1\right) + B/A_0}.
\end{equation}
It is apparent that the both elastic coefficients, $A(=3\gamma A_0)$ and $B$, are important for determination of isothermal compressibility at $T=0$. Since $B \gg A_0$, where $A_0$ is the zero-point energy of Einstein oscillators, it is obvious that for this compressibility an elastic energy plays a dominant role. Finally, let us note that the equation of state derived by Vinet et al. \cite{vinet} for $T=0$ reduces for small $p$ to the form: $\varepsilon \approx \left( 1/B_0 \right) p$, which agrees with our equation(\ref{eq40}) in the limit $A_0 \to 0$.

\subsection{A note on the Gr\"uneisen equation}

The Gr\"uneisen parameter $\gamma$ is introduced by the definition \cite{kittel, girifalco}:
\begin{equation}
\label{eq45}
\gamma \stackrel{def}{=} -\frac{V}{\omega}\,\frac{d \omega}{d V}.
\end{equation}
which is equivalent to our relation (\ref{eq15}).
The typical experimental values of the Gr\"uneisen parameter belong to the range $1 \le \gamma \le 3$ \cite{girifalco}. It can be mentioned that some particular definitions of that parameter, depending on the model, have been discussed in Ref.~\cite{gilvarry}. The experimental determination of $\gamma$ is connected with the so-called Gr\"uneisen equation:
\begin{equation}
\label{eq46}
\frac{\alpha_p}{\kappa_T} = \gamma \, \frac{C_V}{V}
\end{equation}
which has  been tested extensively, especially in the low temperature region. It is worth noticing that the derivation of (\ref{eq46}) can be based on the exact thermodynamic identity:
\begin{equation}
\label{eq47}
\frac{\alpha_p}{\kappa_T} = \left(\frac{\partial p}{\partial T}\right)_V.
\end{equation}
In the case in hand, in order to calculate $\left(\partial p / \partial T \right)_V$, the equation of state $p=p(T,V)$ should be taken in the form of (\ref{eq23}). It can easily be checked that after the calculation of the derivative in (\ref{eq47}), eq.(\ref{eq46}) is satisfied. It should also be noted that the calculations of all quantities in (\ref{eq46}) using the present method require prior solution of the equation of state (\ref{eq29}).
In particular, for the low-temperature region, where $C_V$ tends to zero according to the 3rd law of thermodynamics, we obtain from eq.(\ref{eq46}) that $\alpha_p \to 0$.  This limit is in agreement with the previous result (\ref{eq43}).

%section 3
\section{Numerical results and discussion}

\subsection{Exemplary calculations for the model}

In order to perform the numerical calculations, based on the formulas from the preceding section, it is necessary to estimate the energy constants, from which $A_0$ and $B$ are the most important. The volume elastic modulus in the ground state, $B$, can be found, for instance, from the sound velocity measurements and its experimental value is of the order ($10^{-19} \div 10^{-18}$) J. In turn, $A_0$-coefficient is given by Eq.(\ref{eq26}), where $\Theta_0$ is the Einstein temperature. Assuming that $\Theta_0$ is typically of the order $\sim \,( 10^2 \div 10^3 )$ K we can estimate $A_0$ as $ \sim \,(10^{-21} \div 10^{-20})$ J. Thus, the $A$-constant $(A=3\gamma A_0)$ is about 2 orders smaller than the elastic bulk modulus $B$, and the linear anharmonic term in the Hamiltonian (\ref{eq2}) can be considered as a small correction to the harmonic one. This confirms the fact that such quantities as the thermal expansion coefficient $\alpha_p$ (Eq. \ref{eq38}) or the isothermal compressibility $\kappa_T$ (Eq. \ref{eq39}) are mainly determined by $B$-constant, not by $A_0$. The above estimations of $A$ and $B$ constants, together with the assumed $C=D=0$, yield from Eq.(\ref{eq38}) $\alpha_p \sim \,( 10^{-5} \div 10^{-4} )$ 1/K for the temperatures near the Einstein temperature, which is a physically reasonable order of value. In the same token  assuming a realistic amount of volume per atom, i.e., $V_0/N \sim \,( 10^{-29} \div 10^{-28} )\, {\mathrm m^3}$, we obtain on the basis of Eq.(\ref{eq39}) $\kappa_T \sim \,( 10^{-11} \div 10^{-10} )$ 1/Pa, which is also a correct order for the isothermal compressibility.\\

Although the anharmonic elastic energy $\propto A$ is a small correction in comparison to the harmonic one, it can be of the same order of magnitude as the vibrational energy resulting from Einstein oscillators. Therefore, this anharmonic term is vital in balancing the expanding pressure resulting from oscillators, in the case when the frequency is volume dependent (\ref{eq16}). The expanding pressure arises owing to the decrease of the energy (frequency) of oscillators when they experience collective excitations in increasing volume. The total equilibrium of the system requires balancing of the internal (stretching) forces resulting from expanding quantum oscillators, internal compressive forces of the elastic medium and the external pressure $p$. The resulting deformation $\varepsilon$ is temperature and pressure dependent and can be calculated from the equation of state (\ref{eq29}).
The other energy constants, $C$ and $D$, play merely a correcting role in modeling the static potential, and are important for high temperatures where the elastic deformation $\varepsilon$ is significant.
The exemplary numerical results, presented in this section in Figs.1-8, are obtained for the following set of parameters: $B/A_0=10^2$, $D=0$ and $q=0.74$.\\ 

\begin{figure}
\includegraphics[scale=0.75]{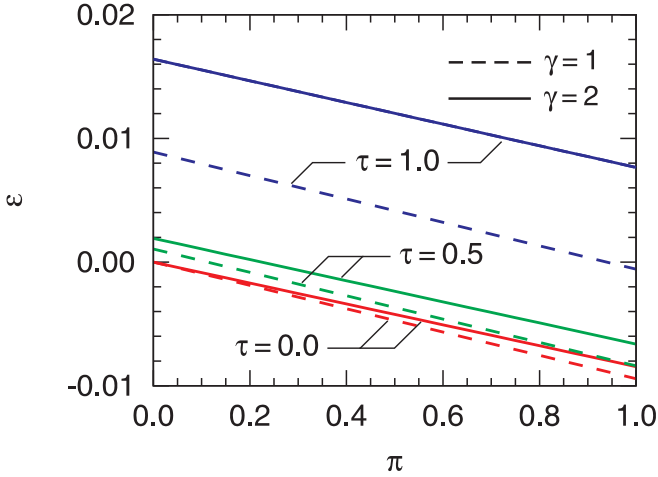}
\caption{Elastic deformation $\varepsilon$ vs. dimensionless pressure $\pi$. The curves are selected for three dimensionless temperatures: $\tau=0$, 0.5 and 1 and two Gr\"uneisen parameters $\gamma=1$ (dashed) and $\gamma=2$ (solid). The elastic energy parameters are : $B/A_0=100$, $C/A_0=D/A_0=0$.}
\label{fig:fig1}
\end{figure}

In Fig.1 we present the dependence of $\varepsilon$ upon $\pi$ for different reduced temperatures $\tau=$ 0, 0.5 and 1, as well as for two selected Gr\"uneisen parameters: $\gamma=1$ and $\gamma=2$. The calculations are based on the equation of state (\ref{eq29}) for $C/A_0=D/A_0=0$. It is seen that these dependencies are almost linear in character, and the slopes of the curves correspond to the isothermal compressibility. It can be deduced from Fig.1 that the isothermal compressibility coefficient should be weakly dependent on pressure in a wide range of temperatures. This observation is in accordance with the analogous assumption in the Murnaghan theory of the equation of state \cite{wallace1}. At higher temperatures (for $\tau=1$) the Gr\"uneisen parameter has remarkable influence on the magnitude of elastic deformation, leading to the increase of $\varepsilon$ when $\gamma$ increases.\\

\begin{figure}
\includegraphics[scale=0.75]{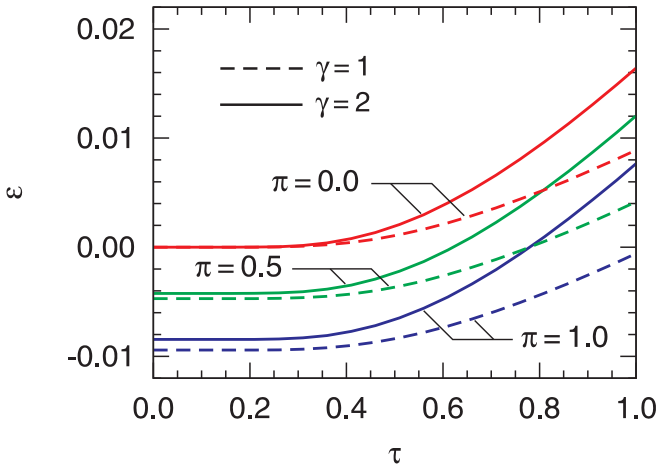}
\caption{Elastic deformation $\varepsilon$ vs. dimensionless temperature $\tau$. The curves are selected for three dimensionless pressures: $\pi=0$, 0.5 and 1 and two Gr\"uneisen parameters $\gamma=1$ (dashed) and $\gamma=2$ (solid). The elastic energy parameters are : $B/A_0=100$, $C/A_0=D/A_0=0$.}
\label{fig:fig2}
\end{figure}

In Fig.2 the dependence of $\varepsilon$ upon $\tau$ for various external pressures ($\pi=$ 0, 0.5 and 1) is presented. The rest of parameters are the same as in Fig.1. In this case the dependencies are not linear and their local slopes are attributed to the thermal expansion coefficient. It is remarkable that by increasing  the Gr\"uneisen parameter $\gamma$ the relative deformation $\varepsilon$ increases, which is evidently pronounced at high temperatures. Such behaviour is in agreement with the previous figure (Fig.1). The dependence presented in this figure is qualitatively similar to the lattice constant dependence on temperature, as can be seen in the Ref.~\cite{kroncke} for the case of AlN. The effect of external pressure on the temperature dependence of relative deformation in Fig.2 resembles, for example, the results obtained in the Ref.~\cite{hama}.\\

\begin{figure}
\includegraphics[scale=0.75]{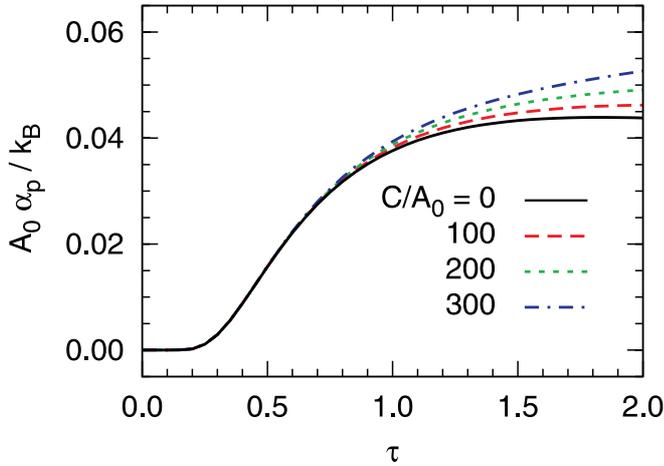}
\caption{The dimensionless thermal expansion coefficient 
$\frac{A_0}{k_{\mathrm B}}\, \alpha_p=\frac {1}{1+\varepsilon}\,\left(\frac{\partial \varepsilon}{\partial \tau}\right)_p$ vs. dimensionless temperature $\tau$. Different curves correspond to $C/A_0=0$, 100, 200 and 300, whereas $B/A_0=100$ and $D/A_0=0$. The external pressure is $\pi=0$ and Gr\"uneisen parameter is $\gamma=2$.}
\label{fig:fig3}
\end{figure}

In Fig.3 the dimensionless thermal expansion coefficient (see Eq.\ref{eq38}) is plotted vs. temperature $\tau$ for external pressure $\pi=0$ and Gr\"uneisen parameter $\gamma=2$. Different curves correspond to various $C/A_0$ parameters. It is seen that $C/A_0$ has influence mainly at high temperatures, where it causes the increase of the thermal expansion coefficient. Let us note, on the basis of Eq. (\ref{eq26}) and (\ref{eq28}), that $\tau=2$ corresponds to the Einstein temperature. In the low-temperature region the thermal expansion coefficient tends to zero independently on $C/A_0$, which is in agreement with the Gr\"uneisen equation (\ref{eq46}). A qualitatively similar behaviour for the thermal expansivity has been found in Ref.~\cite{ponkratz} for the case of MnO, in Ref.~\cite{wallace1} for Ti, Al, NaCl and Na, or in Ref.~\cite{kroncke} for AlN. The conclusion that the anharmonic term becomes important only in the high-temperature region is in agreement with the free energy calculations from first principles \cite{wu}. The similar conclusion can be drawn from the paper Ref.~\cite{oganov} where the influence of intrinsic anharmonicity on the thermodynamic functions has been studied.\\

\begin{figure}
\includegraphics[scale=0.75]{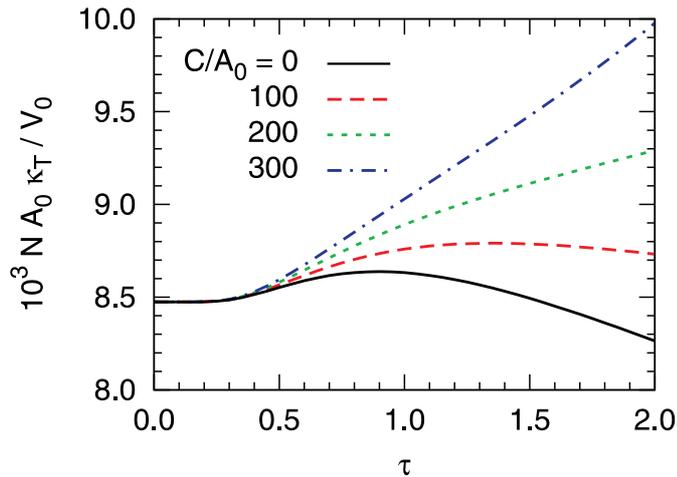}
\caption{The dimensionless isothermal compressibility coefficient $\frac{N A_0}{V_0}\, \kappa_T=-\frac {1}{1+\varepsilon}\,\left(\frac{\partial \varepsilon}{\partial \pi}\right)_T$ vs. dimensionless temperature $\tau$. The parameters are the same as in Fig.3.}
\label{fig:fig4}
\end{figure}

The dimensionless isothermal compressibility (see Eq.\ref{eq39}) is plotted in Fig.4 vs. $\tau$ for the external pressure $\pi=0$ and Gr\"uneisen parameter $\gamma=2$. As in Fig.3, different curves correspond to various $C/A_0$ parameters. A remarkable influence of $C/A_0$-values on the compressibility is seen for the temperatures $\tau >1/2$. On the other hand, for $\tau \to 0$ the compressibility tends to some non-zero value, $\kappa_0$, being in agreement with Eq.(\ref{eq44}). A qualitatively similar experimental results have been obtained for Pb and NaCl. \cite{wallace1} The dependence of isothermal compressibility on temperature corresponds to the fact that the elastic constants are temperature dependent (the exemplary studies can be found in Refs.~\cite{kwon, kim}).\\

\begin{figure}
\includegraphics[scale=0.75]{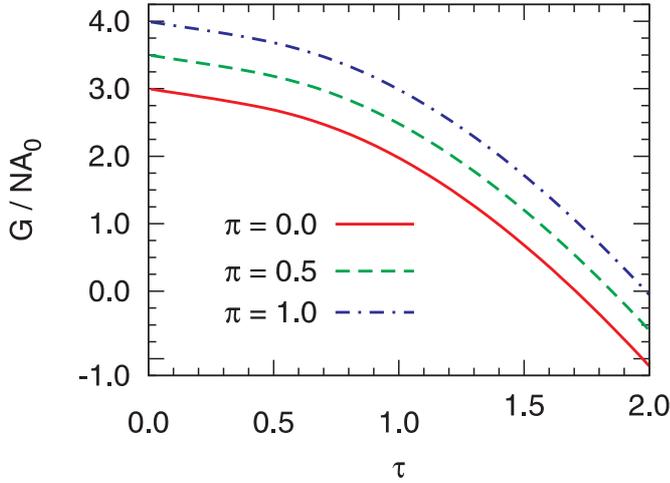}
\caption{The Gibbs free-energy per 1 lattice site in $A_0$ units upon dimensionless temperature $\tau$. Three values of the dimensionless pressure $\pi=0$, 0.5 and 1 are assumed. The Gr\"uneisen parameter is $\gamma=2$ and the elastic energy parameters are: $B/A_0=100$, $C/A_0=D/A_0=0$.}
\label{fig:fig5}
\end{figure}

In Fig.5 the Gibbs free-energy per 1 atom is presented in $A_0$-units vs. reduced temperature $\tau$. The three curves in Fig.5 correspond to different external pressures: $\pi=$ 0, 0.5, and 1, and are plotted by the solid, dashed and dashed-dotted lines, respectively. In this case $C/A_0=0$ and $\gamma=2$. We see that application of the external pressure causes the increase of the Gibbs energy. It is demonstrated in Fig.5 that the Gibbs energy is a concave function of temperature with a monotonously decrease  vs. $\tau$. Such  behaviour indicates thermodynamically stable solution when the entropy (defined by Eq.\ref{eq31}) is positive. Let us note that the initial slope of the Gibbs energy, at $\tau \to 0$ corresponds to the residual entropy.\\

\begin{figure}
\includegraphics[scale=0.75]{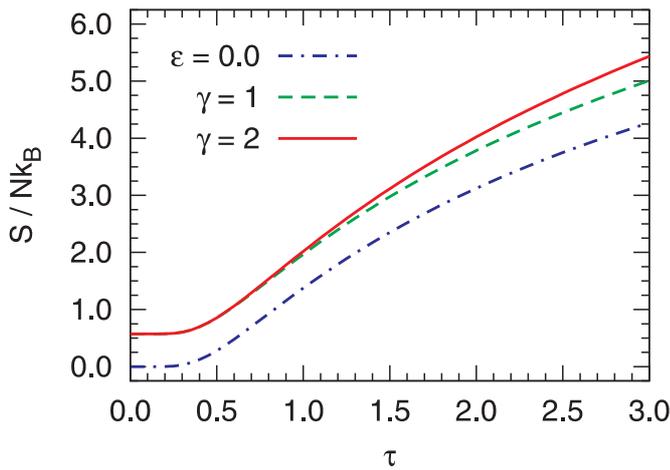}
\caption{Entropy per 1 lattice site in Boltzmann constant units upon dimensionless temperature $\tau$. The dashed and solid lines correspond to $\gamma=1$ and $\gamma=2$, respectively, whereas $B/A_0=100$, $C/A_0=D/A_0=0$ and $\pi=0$. The dashed-dotted line is plotted under constraint $\varepsilon=0$, i.e., for the ideal Einstein model.}
\label{fig:fig6}
\end{figure}

\begin{figure}
\includegraphics[scale=0.75]{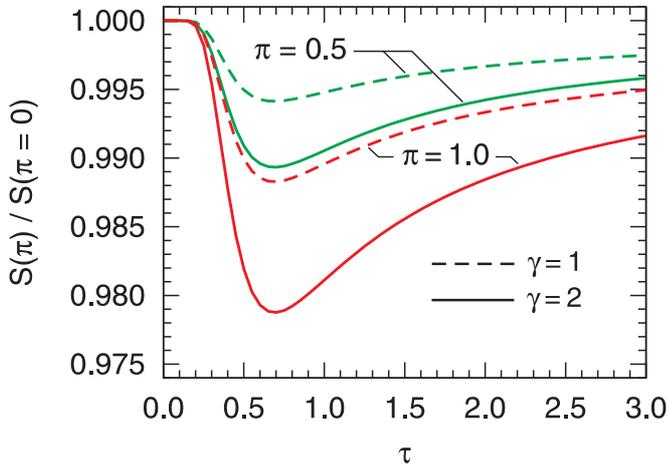}
\caption{Entropy for the dimensionless pressure $\pi$ ($\pi=0.5$ and 1) normalized to the entropy for $\pi=0$, vs. dimensionless temperature $\tau$. The dashed and solid lines correspond to $\gamma=1$ and $\gamma=2$, respectively. The elastic energy parameters are the same as in Fig.6.}
\label{fig:fig7}
\end{figure}

The entropy is illustrated in Figs.6 and 7 as a function of temperature $\tau$. In both figures $C/A_0=0$. In Fig.6 we plotted entropy per 1 lattice site in the Boltzmann constant units, for external pressure $\pi=0$. The solid and dashed curves, which are for $\gamma=1$ and $\gamma=2$, respectively, correspond to our model, where $\varepsilon$ is a function of temperature and is calculated from the equation of state (\ref{eq29}). Due to our choice of integration constant ($c=0$ in Eq.(\ref{eq6})) we see that the residual (configurational) entropy is present, which does not depend on the Gr\"uneisen parameter $\gamma$. Moreover, one can notice that for $T \to0$ the entropy change vs. temperature tend to zero ($\Delta S \to 0$), which is in accordance with the third law of thermodynamics. On the other hand, the increase of $\gamma$ causes some small increase of entropy at high temperatures. For comparison, the dashed-dotted curve presents the entropy for the pure Einstein model, where $\varepsilon$ is put equal to 0 for all temperatures. As stated before, the condition $\varepsilon=0$ is equivalent to the assumption that the frequency of oscillators is kept constant and does not depend on volume. For the pure Einstein model there is no residual entropy in the ground state, since the configurational ordering of atoms is not taken into account.\\ 

\begin{figure}
\includegraphics[scale=0.75]{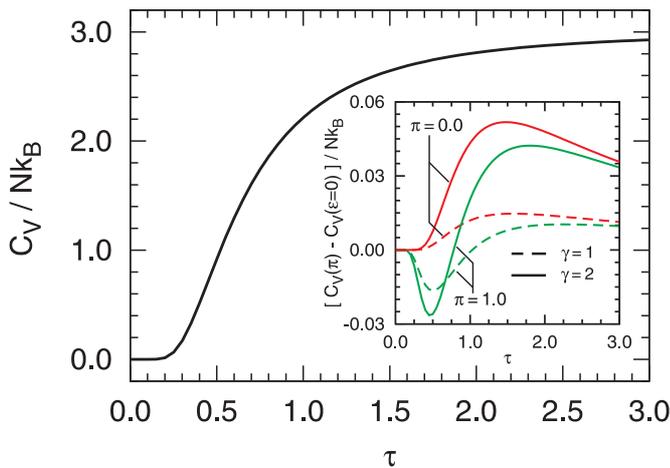}
\caption{Specific heat $C_V$ at constant $V$ per 1 lattice site in $k_{\mathrm B}$ units vs. dimensionless temperature $\tau$. The elastic energy parameters are the same as in Figs. 6 and 7. In the main plot the parameters $\pi=0$ and $\gamma=2$ are assumed. In the inset the difference $C_V\left(\pi\right)-C_V\left(\varepsilon =0\right)$ is plotted for $\pi=0$ and $\pi=1$, as well as for $\gamma=1$ (dashed line) and $\gamma=2$ (solid line).}
\label{fig:fig8}
\end{figure}

It turned out that the entropy is very weakly sensitive to the external pressure, and in Fig.7 we want to study this property in more detail. The relative changes of the entropy $S(\pi)/S(\pi=0)$ are plotted vs. $\tau$ for $\pi= 0.5$ and 1. The dashed and solid lines are for $\gamma=1$ and $\gamma=2$, respectively. We note that the sensitivity of $S$ on $\pi$ is greater for higher Gr\"uneisen parameter $\gamma$. It is also seen that the influence of high external pressure on the ratio $S(\pi)/S(\pi=0)$ is most remarkable for some intermediate temperatures $(\tau \approx 0.6)$, whereas for $\tau \to 0$ the influence of $\pi$ becomes negligible. Again, the behaviour of $S$ near $T=0$ is in agreement with the third law of thermodynamics. The lowering of entropy for $T>0$, when the external pressure is applied, corresponds to the positive thermal expansion coefficient, in accordance with the Maxwell relation: $\left(\partial S / \partial p \right)_T = -\left(\partial V / \partial T \right)_p$.\\

The specific heat per 1 lattice site in the Boltzmann constant units and at constant volume is plotted vs. temperature $\tau$ in Fig.8. The anharmonic parameters are $C/A_0=D/A_0=0$. 
The main plot has been obtained for $\pi=0$ and $\gamma=2$. We note that
the value of $\tau=2$ corresponds to the Einstein temperature. The low-temperature behaviour of the specific heat is typical for the Einstein oscillators, being in agreement with the 3rd law of thermodynamics. On the other hand, the high-temperature part of the curve is classical.
In order to see the difference between the present model and the classical Einstein result in more detail, in the inset, we plot this difference for $\pi=0$ and $\pi=1$, as well as for two Gr\"uneisen parameters: $\gamma=1$ (dashed line) and $\gamma=2$ (solid). $C_V(\pi)$ corresponds to the specific heat of the present model, whereas $C_V(\varepsilon=0)$ is the Einstein result, when $\omega$ does not depend on $V$. It is worth noticing that for $\pi=0$ some small increase of the specific heat occurs for temperatures $T>0$. Similarly to the entropy, the specific heat is only weakly sensitive to the external pressure $\pi$. In particular, one can see that the pressure of $\pi=1$ causes a small decrease of $C_V$ at some restricted range of low temperatures. When the Gr\"uneisen parameter increases all the above changes become enhanced.

\subsection{A comparison with experimental data for solid argon}

For solid argon, which forms FCC structure below the melting temperature of 84 K, the Debye temperature is $\Theta_D$=85 K. From the approximate formula $\Theta_0/\Theta_D=\left(\pi/6\right)^{1/3}$, which can be derived on the basis of Ref.~\cite{kittel}, the Einstein temperature, $\Theta_0$=68.51 K, can be estimated. Hence, on the basis of eq.(\ref{eq26}) we find that the $A_0$-constant amounts to $A_0=0.04729\times10^{-20}$ J. On the other hand, the isothermal compressibility at zero temperature, $\kappa_0$, can be found from the experiment \cite{dobbs}, and for polycrystalline samples it amounts to $\kappa_0=4.1\times10^{-10}\rm {Pa}^{-1}$. The volume per  atom can be estimated as $V_0/N=36.383\times 10^{-30}\, \rm{m}^3$. The Gr\"uneisen parameter can be taken as a mean value of the experimental data from various temperature ranges, \cite{dobbs} which yields $\gamma=2.5$. Having obtained the data given above, on the basis of eq.(\ref{eq44}) the ratio $B/A_0$ can be estimated, yielding $B/A_0=160$. The other elastic energy constants, which are of higher order, i.e., $C$ and $D$, can be treated as the theoretical fitting parameters. We found that the optimal fit to the experimental data (taken from table III, page 561 of Ref.~\cite{dobbs}) is obtained for $C/A_0$=1250 and $D/A_0$=5000, valid for $\epsilon>0$. Taking into account the above set of parameters, we have calculated $\varepsilon,\,\kappa_T,\,\alpha_p$ and $C_V$ for solid argon in the full temperature range $0 < T < 84$ K. To complete the comparison, we supplemented the results with two example isotherms. Since the isotherms cover the range of both positive and negative deformation $\varepsilon$, we decided to allow some asymmetry in our model elastic potential. Thus, for $\varepsilon<0$ we adopted the anharmonic parameter value of $C/A_0$=3000, i.e. the potential is more steep in this range. The results are presented in Figs.9-13 by the solid lines, whereas the experimental data \cite{dobbs} are shown by the open symbols. In addition, the figures have been supplemented with the smoothed experimental data from the Refs.~\cite{peterson,anderson3,tilford}.\\

\begin{figure}
\includegraphics[scale=0.75]{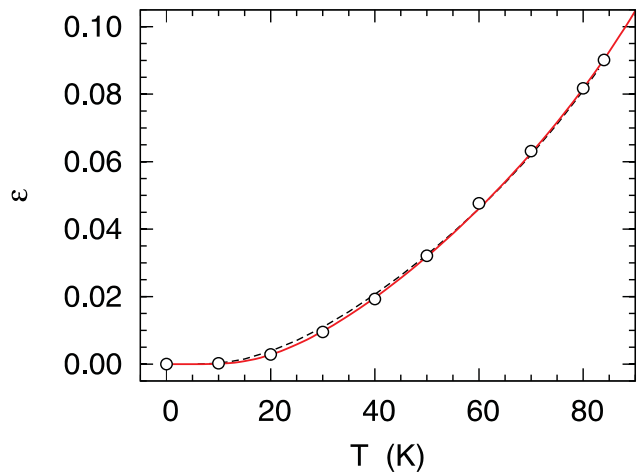}
\caption{Elastic deformation $\varepsilon$ vs. absolute temperature $T$ for solid argon. Solid line - our calculation, dashed line - smoothed experimental data after Ref.~\cite{peterson}, the open symbols - experimental data after Ref.~\cite{dobbs}.}
\label{fig:fig9}
\end{figure}

\begin{figure}
\includegraphics[scale=0.75]{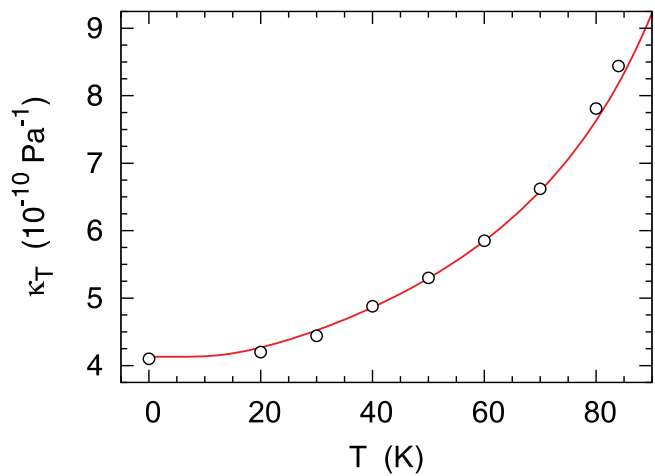}
\caption{Isothermal compressibility coefficient $\kappa_T$ vs. absolute temperature $T$ for solid argon. Solid line - our calculation, the open symbols - experimental data after Ref.~\cite{dobbs}.}
\label{fig:fig10}
\end{figure}

\begin{figure}
\includegraphics[scale=0.75]{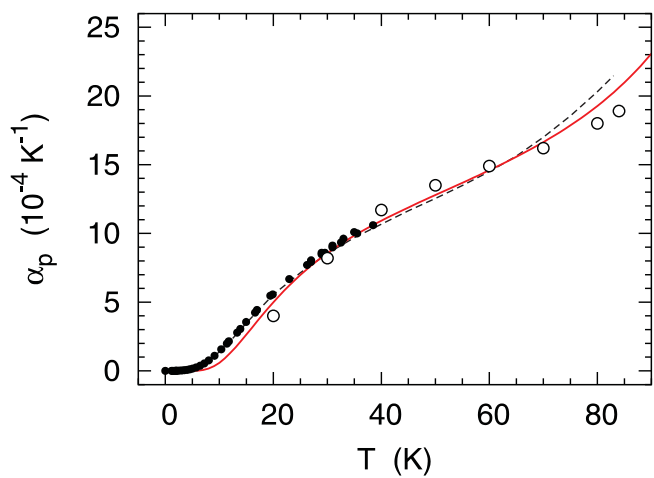}
\caption{Volume thermal expansion coefficient $\alpha_p$ at constant pressure $p=0$ vs. absolute temperature $T$ for solid argon. Solid line - our calculation, dashed line - smoothed experimental data after Ref.~\cite{peterson}, the open symbols - experimental data after Ref.~\cite{dobbs}, the filled symbols - experimental data after Ref.~\cite{tilford}.}
\label{fig:fig11}
\end{figure}

\begin{figure}
\includegraphics[scale=0.75]{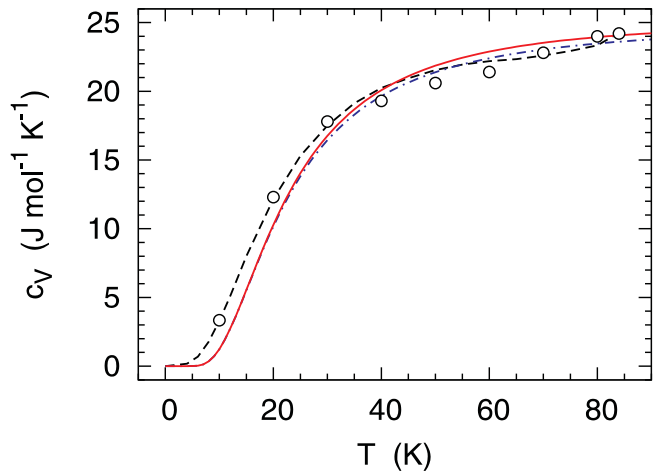}
\caption{Molar specific heat $C_V$ at constant $V$ vs. absolute temperature $T$ for solid argon. Solid line - our calculation, dashed line - smoothed experimental data after Ref.~\cite{peterson}, the open symbols - experimental data after Ref.~\cite{dobbs}. The dashed-dotted line correspond to the calculation with constraint $\varepsilon=0$, i.e., for the pure Einstein model.}
\label{fig:fig12}
\end{figure}

\begin{figure}
\includegraphics[scale=0.75]{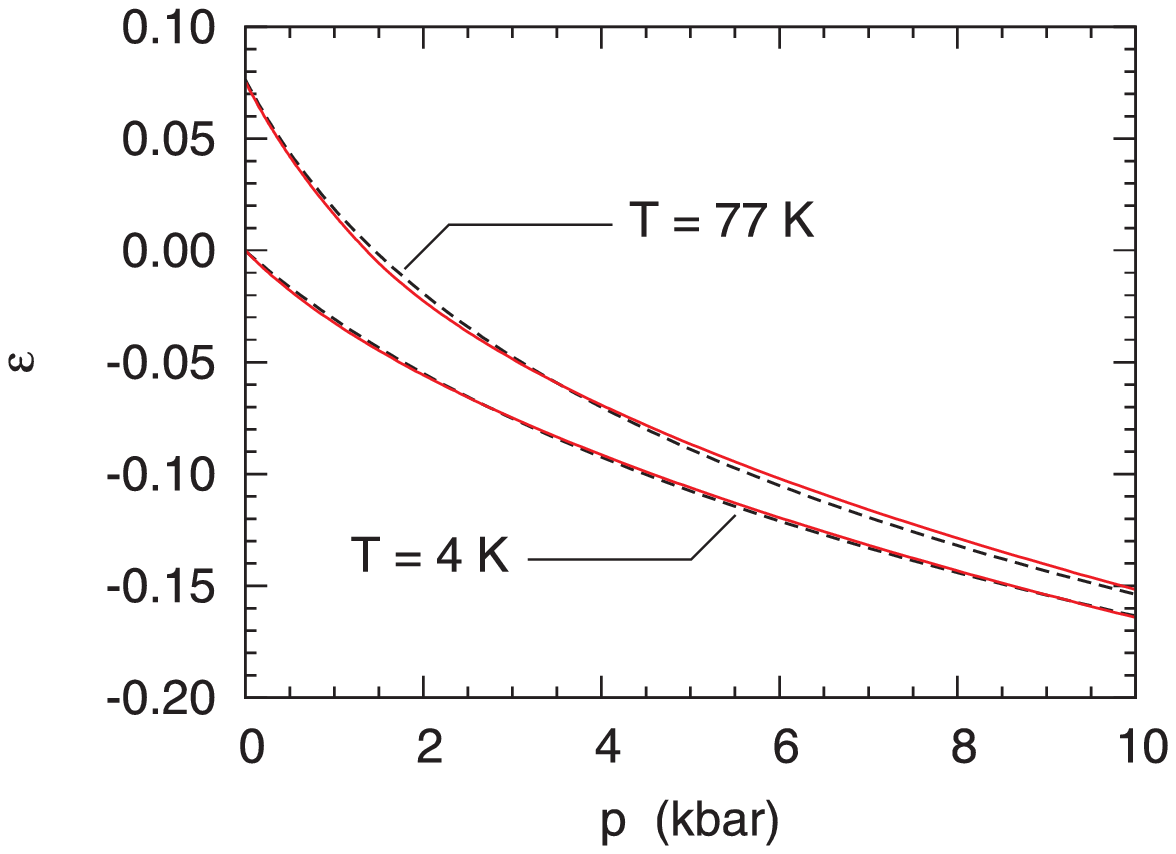}
\caption{Isotherms for solid argon (dependence of elastic deformation on external pressure) for two selected temperatures. Solid lines - our calculation, dashed lines - smoothed experimental data after Ref.~\cite{anderson3}.}
\label{fig:fig13}
\end{figure}

In Fig.9 the calculation of elastic deformation $\varepsilon$ vs. absolute temperature $T$ is presented. 
The derivative of elastic deformation over the pressure, i.e., the isothermal compressibility coefficient, $\kappa_T$, is plotted vs. $T$ in Fig.10.  It can be noted that in both figures (9 and 10) the experimental dependencies  are nonlinear, in agreement with the present theory.
In turn, the thermal expansion coefficient, $\alpha_p$, at constant pressure ($p=0$) is plotted in Fig.11 for the same temperature range as in Figs. 9 and 10. 
Finally, the molar specific heat $C_V$ at constant $V$ is presented in Fig.12. For comparison, in Fig.12 the dashed line is plotted for the pure Einstein model when we impose the constraint $\varepsilon=0$.\\

Our calculated isothermal compressibility, $\kappa_T$, can be related to the adiabatic compressibility, $\kappa_S$, via the formula:
\begin{equation}
\label{eq50}
\kappa_T = \kappa_S\left(1+ \alpha_p \gamma T\right).
\end{equation}
It can be checked that the above formula is equivalent to the Gr\"uneisen equation (\ref{eq46}) which is also satisfied in our case. As pointed out in Ref.~\cite{dobbs}, the values of 
$\kappa_S$ and $\kappa_T$ can be measured independently; $\kappa_S$ by the ultrasonic method from isentropic sound velocity, and $\kappa_T$ by the piston-displacement method of Bridgman. Using the formula (\ref{eq50}) a satisfactory agreement between those two measurements has been obtained in Ref.~\cite{dobbs}. Such consistency directly transfers to our calculations, therefore in Fig.10 only $\kappa_T$ curve has been presented.

Regarding the isothermal compressibility  $\kappa_0$ (at $p=0$, $T=0$) we adopted here the value $\kappa_0=4.1\times10^{-10}~\rm {Pa}^{-1}$ for polycrystalline samples after Dobbs and Jones \cite{dobbs}. On this basis the classical (atomic volume) bulk modulus, $B^{\prime}$, can be estimated at $T=0$ K:  $B^{\prime}=1/\kappa_0=0.244\times 10^{10}~\rm {N/m^2}$. On the other hand, for single cubic crystals the bulk modulus can be obtained from the formula \cite{kittel} $B^{\prime}=\left(C_{11}+2C_{12}\right)/3$, if the elastic constants $C_{11}$ and $C_{12}$ are known. These constants for argon single crystals have been deduced from measured isentropic sound velocity \cite{keeler}. The values extrapolated to $T=0$ K are: $C_{11}=4.39$ and $C_{12}=1.83$ in units of $10^{9}~\rm {N/m^2}$ \cite{keeler}. Hence, the bulk modulus $B^{\prime}$ obtained on this basis is $B^{\prime}=0.268\times10^{10}~\rm {N/m^2}$. It can be concluded that this value of $B^{\prime}$ for single crystal is about $10\%$ larger than for polycrystalline samples given in \cite{dobbs}.

In the Fig.~13 two selected isotherms are shown (low- and high-temperature one), presenting elastic deformation $\varepsilon$ as a function of external pressure $p$. It is visible that using a model asymmetric potential, a good fit to experimental data in whole pressure range studied is obtained, both for high- and low-temperature results. This emphasizes the pronounced importance of the exact form of the static lattice potential for pressure studies, since the deformations here exceed 10\%. The selected potential provides reliable description for pressures up to 10 kbar. In light of the assumption that our static elastic potential originates from the expansion at the point $T=0$ and $p=0$, further adjusting the potential would improve the consistency between calculations and experimental data for higher pressures.

It can be concluded, on the basis of Figs. 9-13, that the agreement between calculated lines and the experimental data is quite satisfactory. One should take into account that all these curves have been calculated selfconsistently for the same set of parameters, as described above in this subsection. 
It can be supposed that the fitting could even be better if we would allow the Gr\"uneisen parameter to vary with temperature and pressure, as it has been suggested from the experimental measurements \cite{dobbs,ross}.
As far as the specific heat is concerned, the new result (solid line) differs only insignificantly from the Einstein result (dashed), and the difference is mainly noteworthy at high temperatures. However, it should be noted that the dashed line in Fig.12 is the only result of the pure Einstein model which can be compared with all these experimental data. The full description of temperature dependencies of remaining quantities, such as $\varepsilon,\, \kappa_T$ and $\alpha_p$ has been possible in the improved approach to the Einstein model, when the equation of state (\ref{eq29}) is taken into account.

\subsection{Final remarks}

In the present paper a simple combination of the Einstein and elastic models of solid state is presented. First of all, the dependency of frequency of quantum oscillators on the volume has been introduced in a simplified way via Gr\"uneisen assumption. Simultaneously, the elastic properties of the crystal have been taken into account via classical Hamiltonian, containing anharmonic terms.\\

An idea that the free-energy is a sum of several components has been presented, for instance, in the book of Wallace \cite{wallace2}. However, we have shown that exploitation of this idea led us to the new form of the equation of state (\ref{eq29}). Contrary to the Birch-Murnaghan equation of state, which presents only isothermal description \cite{wallace2}, in our equation the temperature $T$ plays a role equivalent to the rest of variables, i.e., $p$ and $V$.
In particular, the thermal expansion coefficient, as well as the isothermal compressibility have simultaneously been derived from this new equation of state. It is well-known that these quantities could not be inferred from the sole Einstein model. On the other hand, our equation of state includes also the pressure resulting from expanding quantum oscillators. Moreover, comparing our method with the papers based on the Wallace approach, one should notice that one of our anharmonic parameters in the static potential, namely $A$, is not independent, but has been related to the Einstein temperature (via Eqs.~\ref{eq27} and \ref{eq28}). As we pointed out, such relationship assures that the equilibrium condition for the total free energy at $p=0$ and $T=0$ is satisfied and the system of quantum oscillators becomes stable. When this point is not discussed, the functional form of the static lattice energy can also be adopted in the form given in Ref.~\cite{greeff} (after Vinet et al.). That form is much more complicated than our polynomial approach, nevertheless it has successfully been applied for very high pressures.\\

 The present method requires several constant parameters, such as: the Einstein temperature, isothermal compressibility (or the volume elastic modulus) at the absolute zero temperature, as well as the Gr\"uneisen parameter, which should be taken from experiment.\\

The region of applicability of the model (its temperature and pressure range of validity) is mainly connected with the number of terms which are taken into account in the polynomial form of elastic Hamiltonian (\ref{eq2}). This form is based on the assumption that the equilibrium central point is $\varepsilon = 0$ for $T=0$ and $p=0$. Thus, the theory is best applicable for the expansion around this equilibrium. However, as we have seen on the example of solid argon, merely first four terms were sufficient to describe deformation $\varepsilon$ in a wide range; starting from $\varepsilon \approx 0.1$ (near the melting temperature and $p=0$, Fig.9) down to $\varepsilon \approx -0.16$ (for $T=4$ K and $p=10$ kbar, Fig.13). Of course, validity of the model for lower values of $\varepsilon < 0$ (for very high pressure, where anharmonicity of static potential plays a role) would require higher order terms in (\ref{eq2}). One should also remember that our assumption concerning the space filling coefficient ($q= const.$) would not be fulfilled at extremely high pressures. As far as the vibrational anharmonic effects are considered, they all are taken into account by the effective Gr\"uneisen parameter $\gamma$. We think that the approach is useful in the range of temperatures up to the melting point.
In conclusion, the presented formulation allows for a complete thermodynamic description of the system in a relatively wide range of external pressure and temperature.
Our considerations are related to the quasistatic processes. Therefore, the shock-wave experiments cannot be described within this model.\\

It should be noted that the method is relatively simple, gives analytical form of the equation of state, and therefore it can serve as a first approximation for more advanced approaches.
For instance, in the prototype calculations we have used only single variable $\varepsilon$ for description of the volume elastic deformation. However, it seems possible to generalize the approach for anisotropic deformations and anisotropic external pressures, involving also the Poisson coefficient.
The presented method can be potentially extended basing on the Debye approximation, in which the linear dispersion relation for collective excitations together with the proper density of states are taken into consideration. For instance, in Ref.~\cite{greeff} the vibrational free energy has been taken into account in the high temperature Debye model. In that approach each moment of the density of states function requires a separate Gr\"uneisen parameter, which makes the method much more complicated. On the other hand, for metallic systems the electronic part of the free-energy should also be included \cite{greeff}. However, such extensions of the method need further studies and should be a subject of separate assignment.\\  

%\ack
The paper has been partly supported by the grant VEGA 1/0431/10.

%\section*{References}


\begin{thebibliography}{48}

\expandafter\ifx\csname natexlab\endcsname\relax\def\natexlab#1{#1}\fi
\expandafter\ifx\csname bibnamefont\endcsname\relax
  \def\bibnamefont#1{#1}\fi
\expandafter\ifx\csname bibfnamefont\endcsname\relax
  \def\bibfnamefont#1{#1}\fi
\expandafter\ifx\csname citenamefont\endcsname\relax
  \def\citenamefont#1{#1}\fi
\expandafter\ifx\csname url\endcsname\relax
  \def\url#1{\texttt{#1}}\fi
\expandafter\ifx\csname urlprefix\endcsname\relax\def\urlprefix{URL }\fi
\providecommand{\bibinfo}[2]{#2}
\providecommand{\eprint}[2][]{\url{#2}}
 
\bibitem{einstein}
\bibinfo{author}{\bibfnamefont{A.}~\bibnamefont{Einstein}},
\bibinfo{journal}{Annalen der Physik} \textbf{\bibinfo{volume}{22}},
\bibinfo{pages}{180} (\bibinfo{year}{1907}).
  
\bibitem{kittel}
\bibinfo{author}{\bibfnamefont{C.}~\bibnamefont{Kittel}}, 
\emph{\bibinfo{booktitle}{Introduction to Solid State Physics}},
(\bibinfo{publisher}{John Wiley \& Sons, Inc.},
\bibinfo{year}{1996}).    
  
\bibitem{girifalco}
\bibinfo{author}{\bibfnamefont{L. A.}~\bibnamefont{Girifalco}}, 
\emph{\bibinfo{booktitle}{Statistical Mechanics of Solids}},
(\bibinfo{publisher}{Oxford University Press},
\bibinfo{year}{2000}).      

\bibitem{caplin}
\bibinfo{author}{\bibfnamefont{A. D.}~\bibnamefont{Caplin}},
\bibinfo{author}{\bibfnamefont{G.}~\bibnamefont{Gr\"uner}},
\bibinfo{author}{\bibfnamefont{J. B.}~\bibnamefont{Dunlop}},
\bibinfo{journal}{Phys. Rev. Lett.} \textbf{\bibinfo{volume}{30}},
\bibinfo{pages}{1138} (\bibinfo{year}{1973}).  

\bibitem{avramov}
\bibinfo{author}{\bibfnamefont{I.}~\bibnamefont{Avramov}},
\bibinfo{author}{\bibfnamefont{M.}~\bibnamefont{Michailov}},
\bibinfo{journal}{J. Phys.: Condens. Matter} \textbf{\bibinfo{volume}{20}},
\bibinfo{pages}{295224} (\bibinfo{year}{2008}). 

 
\bibitem{debye}
\bibinfo{author}{\bibfnamefont{P.}~\bibnamefont{Debye}},
\bibinfo{journal}{Annalen der Physik} \textbf{\bibinfo{volume}{39}},
\bibinfo{pages}{789} (\bibinfo{year}{1912}). 

\bibitem{zubov}
\bibinfo{author}{\bibfnamefont{V. I.}~\bibnamefont{Zubov}},
\bibinfo{journal}{Cryst. Res. Technol.} \textbf{\bibinfo{volume}{30}},
\bibinfo{pages}{149} (\bibinfo{year}{1995}). 

\bibitem{li}
\bibinfo{author}{\bibfnamefont{D.}~\bibnamefont{Li}},
\bibinfo{author}{\bibfnamefont{G. A.}~\bibnamefont{Voth}},
\bibinfo{journal}{J. Chem. Phys.} \textbf{\bibinfo{volume}{96}},
\bibinfo{pages}{5340} (\bibinfo{year}{1992}). 

\bibitem{cankurtaran}
\bibinfo{author}{\bibfnamefont{M.}~\bibnamefont{Cankurtaran}},
\bibinfo{author}{\bibfnamefont{B. M.}~\bibnamefont{Askerov}},
\bibinfo{journal}{Phys. Stat. Sol. (b)} \textbf{\bibinfo{volume}{194}},
\bibinfo{pages}{499} (\bibinfo{year}{1996}). 

\bibitem{lusk}
\bibinfo{author}{\bibfnamefont{M. T.}~\bibnamefont{Lusk}},
\bibinfo{journal}{J. Chem. Phys.} \textbf{\bibinfo{volume}{121}},
\bibinfo{pages}{11208} (\bibinfo{year}{2004});
\bibinfo{journal}{Phys. Rev. B} \textbf{\bibinfo{volume}{70}},
\bibinfo{pages}{174103} (\bibinfo{year}{2004}). 

\bibitem{holzapfel}
\bibinfo{author}{\bibfnamefont{W. B.}~\bibnamefont{Holzapfel}},
\bibinfo{author}{\bibfnamefont{M.}~\bibnamefont{Hartwig}},
\bibinfo{author}{\bibfnamefont{W.}~\bibnamefont{Sievers}},
\bibinfo{journal}{J. Phys. Chem. Ref. Data} \textbf{\bibinfo{volume}{30}},
\bibinfo{pages}{515} (\bibinfo{year}{2001}). 

\bibitem{ponkratz}
\bibinfo{author}{\bibfnamefont{U.}~\bibnamefont{Ponkratz}},
\bibinfo{author}{\bibfnamefont{W. B.}~\bibnamefont{Holzapfel}},
\bibinfo{journal}{J. Phys.: Condens. Matter} \textbf{\bibinfo{volume}{16}},
\bibinfo{pages}{S963} (\bibinfo{year}{2004}). 

\bibitem{murnaghan}
\bibinfo{author}{\bibfnamefont{F. D.}~\bibnamefont{Murnaghan}},
\bibinfo{journal}{Amer. J. Math.} \textbf{\bibinfo{volume}{59}},
\bibinfo{pages}{235} (\bibinfo{year}{1937}). 

\bibitem{birch1}
\bibinfo{author}{\bibfnamefont{F.}~\bibnamefont{Birch}},
\bibinfo{journal}{J. Appl. Phys.} \textbf{\bibinfo{volume}{9}},
\bibinfo{pages}{279} (\bibinfo{year}{1938}). 

\bibitem{slater}
\bibinfo{author}{\bibfnamefont{J. C.}~\bibnamefont{Slater}}, 
\emph{\bibinfo{booktitle}{Introduction to Chemical Physics}},
(\bibinfo{publisher}{McGraw-Hill Book Co.},
\bibinfo{year}{1939}).   

\bibitem{bridgman}
\bibinfo{author}{\bibfnamefont{P. W.}~\bibnamefont{Bridgman}}, 
\emph{\bibinfo{booktitle}{The Physics of High Pressure}},
(\bibinfo{publisher}{G. Bell and Sons Ltd.},
\bibinfo{year}{1949}).   

\bibitem{barsch}
\bibinfo{author}{\bibfnamefont{G. R.}~\bibnamefont{Barsch}},
\bibinfo{journal}{J. Appl. Phys.} \textbf{\bibinfo{volume}{39}},
\bibinfo{pages}{3780} (\bibinfo{year}{1968}). 

\bibitem{wallace1}
\bibinfo{author}{\bibfnamefont{D. C.}~\bibnamefont{Wallace}}, 
\emph{\bibinfo{booktitle}{Thermodynamics of Crystals}},
(\bibinfo{publisher}{John Wiley \& Sons, Inc.},
\bibinfo{year}{1972}).    

\bibitem{birch2}
\bibinfo{author}{\bibfnamefont{F.}~\bibnamefont{Birch}},
\bibinfo{journal}{J. Geophys. Res.} \textbf{\bibinfo{volume}{83}},
\bibinfo{pages}{1257} (\bibinfo{year}{1978}). 

\bibitem{vinet2}
\bibinfo{author}{\bibfnamefont{P.}~\bibnamefont{Vinet}},
\bibinfo{author}{\bibfnamefont{J. R.}~\bibnamefont{Smith}},
\bibinfo{author}{\bibfnamefont{J.}~\bibnamefont{Ferrante}},
\bibinfo{author}{\bibfnamefont{J. H.}~\bibnamefont{Rose}},
\bibinfo{journal}{Phys. Rev. B} \textbf{\bibinfo{volume}{35}},
\bibinfo{pages}{1945} (\bibinfo{year}{1987}).

\bibitem{vinet}
\bibinfo{author}{\bibfnamefont{P.}~\bibnamefont{Vinet}},
\bibinfo{author}{\bibfnamefont{J. H.}~\bibnamefont{Rose}},
\bibinfo{author}{\bibfnamefont{J.}~\bibnamefont{Ferrante}},
\bibinfo{author}{\bibfnamefont{J. R.}~\bibnamefont{Smith}},
\bibinfo{journal}{J. Phys.: Condens. Matter} \textbf{\bibinfo{volume}{1}},
\bibinfo{pages}{1941} (\bibinfo{year}{1989}). 

\bibitem{wallace2}
\bibinfo{author}{\bibfnamefont{D. C.}~\bibnamefont{Wallace}}, 
\emph{\bibinfo{booktitle}{Statistical Physics of Crystals and Liquids}},
(\bibinfo{publisher}{World Scientific},
\bibinfo{year}{2002}).  

\bibitem{anderson1}
\bibinfo{author}{\bibfnamefont{M. S.}~\bibnamefont{Anderson}},
\bibinfo{author}{\bibfnamefont{C. A.}~\bibnamefont{Swenson}},
\bibinfo{journal}{Phys. Rev. B} \textbf{\bibinfo{volume}{10}},
\bibinfo{pages}{5184} (\bibinfo{year}{1974}). 

\bibitem{anderson2}
\bibinfo{author}{\bibfnamefont{M. S.}~\bibnamefont{Anderson}},
\bibinfo{author}{\bibfnamefont{C. A.}~\bibnamefont{Swenson}},
\bibinfo{journal}{Phys. Rev. B} \textbf{\bibinfo{volume}{28}},
\bibinfo{pages}{5395} (\bibinfo{year}{1983}). 

\bibitem{roy}
\bibinfo{author}{\bibfnamefont{P. B.}~\bibnamefont{Roy}},
\bibinfo{author}{\bibfnamefont{S. B.}~\bibnamefont{Roy}},
\bibinfo{journal}{J. Phys.: Condens. Matter} \textbf{\bibinfo{volume}{17}},
\bibinfo{pages}{6193} (\bibinfo{year}{2005}). 

\bibitem{hama}
\bibinfo{author}{\bibfnamefont{J.}~\bibnamefont{Hama}},
\bibinfo{author}{\bibfnamefont{K.}~\bibnamefont{Suito}},
\bibinfo{journal}{J. Phys.: Condens. Matter} \textbf{\bibinfo{volume}{8}},
\bibinfo{pages}{67} (\bibinfo{year}{1996}). 

\bibitem{foata}
\bibinfo{author}{\bibfnamefont{M.}~\bibnamefont{Foata-Prestavoine}},
\bibinfo{author}{\bibfnamefont{G.}~\bibnamefont{Robert}},
\bibinfo{author}{\bibfnamefont{M.-H.}~\bibnamefont{Nadal}},
\bibinfo{author}{\bibfnamefont{S.}~\bibnamefont{Bernard}},
\bibinfo{journal}{Phys. Rev. B} \textbf{\bibinfo{volume}{76}},
\bibinfo{pages}{104104} (\bibinfo{year}{2007}). 

\bibitem{wu}
\bibinfo{author}{\bibfnamefont{Z.}~\bibnamefont{Wu}},
\bibinfo{journal}{Phys. Rev. B} \textbf{\bibinfo{volume}{81}},
\bibinfo{pages}{172301} (\bibinfo{year}{2010}). 


\bibitem{chisolm}
\bibinfo{author}{\bibfnamefont{E. D.}~\bibnamefont{Chisolm}},
\bibinfo{author}{\bibfnamefont{S. D.}~\bibnamefont{Crockett}},
\bibinfo{author}{\bibfnamefont{D. C.}~\bibnamefont{Wallace}},
\bibinfo{journal}{Phys. Rev. B} \textbf{\bibinfo{volume}{68}},
\bibinfo{pages}{104103} (\bibinfo{year}{2003}). 

\bibitem{greeff}
\bibinfo{author}{\bibfnamefont{C. W.}~\bibnamefont{Greeff}},
\bibinfo{author}{\bibfnamefont{M. J.}~\bibnamefont{Graf}},
\bibinfo{journal}{Phys. Rev. B} \textbf{\bibinfo{volume}{69}},
\bibinfo{pages}{054107} (\bibinfo{year}{2004}). 

\bibitem{gruneisen}
\bibinfo{author}{\bibfnamefont{E.}~\bibnamefont{Gr\"uneisen}},
\bibinfo{journal}{Annalen der Physik} \textbf{\bibinfo{volume}{39}},
\bibinfo{pages}{257} (\bibinfo{year}{1912});
\emph{\bibinfo{booktitle}{Handbuch der Physik}},
(\bibinfo{publisher}{Verlag Julius Springer},
\bibinfo{address}{Berlin}, \bibinfo{year}{1926}). 

\bibitem{kwon}
\bibinfo{author}{\bibfnamefont{Tai-Hyung}~\bibnamefont{Kwon}},
\bibinfo{journal}{Journal of the Korean Physical Society} \textbf{\bibinfo{volume}{26}},
\bibinfo{pages}{349} (\bibinfo{year}{1993}). 

\bibitem{hill}
\bibinfo{author}{\bibfnamefont{T. L.}~\bibnamefont{Hill}},
\emph{\bibinfo{booktitle}{Statistical Mechanics}},
(\bibinfo{publisher}{McGraw-Hill Book Co.},
\bibinfo{year}{1956}).      

\bibitem{reif}
\bibinfo{author}{\bibfnamefont{F.}~\bibnamefont{Reif}},
\emph{\bibinfo{booktitle}{Fundamentals of Statistical and Thermal Physics}},
(\bibinfo{publisher}{McGraw-Hill Book Co.},
\bibinfo{year}{1985}).      

\bibitem{gilvarry}
\bibinfo{author}{\bibfnamefont{J. J.}~\bibnamefont{Gilvarry}},
\bibinfo{journal}{Phys. Rev.} \textbf{\bibinfo{volume}{102}},
\bibinfo{pages}{331} (\bibinfo{year}{1956}). 

\bibitem{kroncke}
\bibinfo{author}{\bibfnamefont{H.}~\bibnamefont{Kr\"oncke}},
\bibinfo{author}{\bibfnamefont{S.}~\bibnamefont{Figge}},
\bibinfo{author}{\bibfnamefont{B. M.}~\bibnamefont{Epelbaum}},
\bibinfo{author}{\bibfnamefont{D.}~\bibnamefont{Hommel}},
\bibinfo{journal}{Acta Phys. Polon. A} \textbf{\bibinfo{volume}{114}},
\bibinfo{pages}{1193} (\bibinfo{year}{2008}). 

\bibitem{oganov}
\bibinfo{author}{\bibfnamefont{A. R.}~\bibnamefont{Oganov}},
\bibinfo{author}{\bibfnamefont{P. I.}~\bibnamefont{Dorogokupets}},
\bibinfo{journal}{J. Phys.: Condens. Matter} \textbf{\bibinfo{volume}{16}},
\bibinfo{pages}{1351} (\bibinfo{year}{2004}). 


\bibitem{kim}
\bibinfo{author}{\bibfnamefont{S.-C.}~\bibnamefont{Kim}},
\bibinfo{author}{\bibfnamefont{T. H.}~\bibnamefont{Kwon}},
\bibinfo{journal}{Phys. Rev. B} \textbf{\bibinfo{volume}{45}},
\bibinfo{pages}{2105} (\bibinfo{year}{1992}).   

\bibitem{dobbs}
\bibinfo{author}{\bibfnamefont{E. K.}~\bibnamefont{Dobbs}},
\bibinfo{author}{\bibfnamefont{G. O.}~\bibnamefont{Jones}},
\bibinfo{journal}{Rep. Prog. Phys.} \textbf{\bibinfo{volume}{20}},
\bibinfo{pages}{516} (\bibinfo{year}{1957}).   


\bibitem{peterson}
\bibinfo{author}{\bibfnamefont{O. G.}~\bibnamefont{Peterson}},
\bibinfo{author}{\bibfnamefont{D. N.}~\bibnamefont{Batchelder}},
\bibinfo{author}{\bibfnamefont{R. O.}~\bibnamefont{Simmons}},
\bibinfo{journal}{Phys. Rev.} \textbf{\bibinfo{volume}{150}},
\bibinfo{pages}{703} (\bibinfo{year}{1966}).   
 
 \bibitem{anderson3}
\bibinfo{author}{\bibfnamefont{M. S.}~\bibnamefont{Anderson}},
\bibinfo{author}{\bibfnamefont{C. A.}~\bibnamefont{Swenson}},
\bibinfo{journal}{J. Phys. Chem. Solids} \textbf{\bibinfo{volume}{36}},
\bibinfo{pages}{145} (\bibinfo{year}{1975}).  
 

\bibitem{tilford}
\bibinfo{author}{\bibfnamefont{C. R.}~\bibnamefont{Tilford}},
\bibinfo{author}{\bibfnamefont{C. A.}~\bibnamefont{Swenson}},
\bibinfo{journal}{Phys. Rev. B} \textbf{\bibinfo{volume}{5}},
\bibinfo{pages}{719} (\bibinfo{year}{1972}).

\bibitem{keeler}
\bibinfo{author}{\bibfnamefont{G. J.}~\bibnamefont{Keeler}},
\bibinfo{author}{\bibfnamefont{D. N.}~\bibnamefont{Batchelder}},
\bibinfo{journal}{J. Phys. C: Solid State Phys.} \textbf{\bibinfo{volume}{3}},
\bibinfo{pages}{510} (\bibinfo{year}{1970}).

\bibitem{ross}
\bibinfo{author}{\bibfnamefont{M.}~\bibnamefont{Ross}},
\bibinfo{author}{\bibfnamefont{H. K.}~\bibnamefont{Mao}},
\bibinfo{author}{\bibfnamefont{P. M.}~\bibnamefont{Bell}},
\bibinfo{author}{\bibfnamefont{J. A.}~\bibnamefont{Xu}},
\bibinfo{journal}{J. Chem. Phys.} \textbf{\bibinfo{volume}{85}},
\bibinfo{pages}{1028} (\bibinfo{year}{1986}).

%%%







 
\end{thebibliography}
\end{document}